\documentclass[a4paper,twoside]{article}
\usepackage{amsmath,amssymb,amsfonts}
\pdfoutput=1
\usepackage{graphics}

\begin{document}
\title{An inhomogeneous stochastic rate process for evolution
from states in an information geometric neighbourhood of uniform fitness\footnote{Invited paper at 
3$^{rd}$ Conference on Information Geometry and its Applications, 
Max-Planck-Institut f\"{u}r Mathematik in den Naturwissenschaften, Leipzig, 2-6 August 2010.  }}
\author{ C.T.J. Dodson \\{\small\it School of Mathematics, University of Manchester,
  Manchester M13 9PL, UK}\\
  {\small\it ctdodson@manchester.ac.uk}}
\pagestyle{myheadings}
\markboth{Inhomogeneous evolution
from a neighbourhood of uniform fitness}{C.T.J. Dodson}
\date{ }
\maketitle

\begin{abstract}
This study elaborates some examples of a simple evolutionary stochastic rate process
where the population rate of change depends on the distribution of properties---so
different cohorts change at different rates. We investigate
the effect on the evolution arising from parametrized perturbations of
uniformity for the initial inhomogeneity. The information geometric
neighbourhood system yields also solutions
for a wide range of other initial inhomogeneity distributions,
including approximations to truncated Gaussians of arbitrarily small variance
and distributions with pronounced extreme values.
It is found that, under quite
considerable alterations in the shape and variance of the initial distribution of inhomogeneity
in unfitness, the decline of the mean does change markedly with the variation in starting conditions,
but the net population evolution seems surprisingly stable.

\end{abstract}
\section{Introduction}
Consider a population with an inhomogeneous property distribution
faced with a sudden environmental change at $t=0$ and let
 $N(t)$ represent the declining population of more unfit individuals.
In our example the distribution of unfitness $a$ will lie in the range $[0,1]$ and under
selective evolution the fraction $N(t)$ could consist of individuals with unfitness
 below some threshold value which itself may evolve.
However, as $N(t)$ evolves we expect the distribution of unfitness to become
 skewed increasigly towards smaller values, so improving population fitness.
 We make use of existing inhomogeneous stochastic rate
process theory \cite{Karev03,Karev09}, and information geometry,
\cite{AN,InfoGeom,entflow,stochrateproc}.
\begin{figure}
\begin{center}
\begin{picture}(300,150)(0,0)
\put(0,-20){\resizebox{10 cm}{!}{\includegraphics{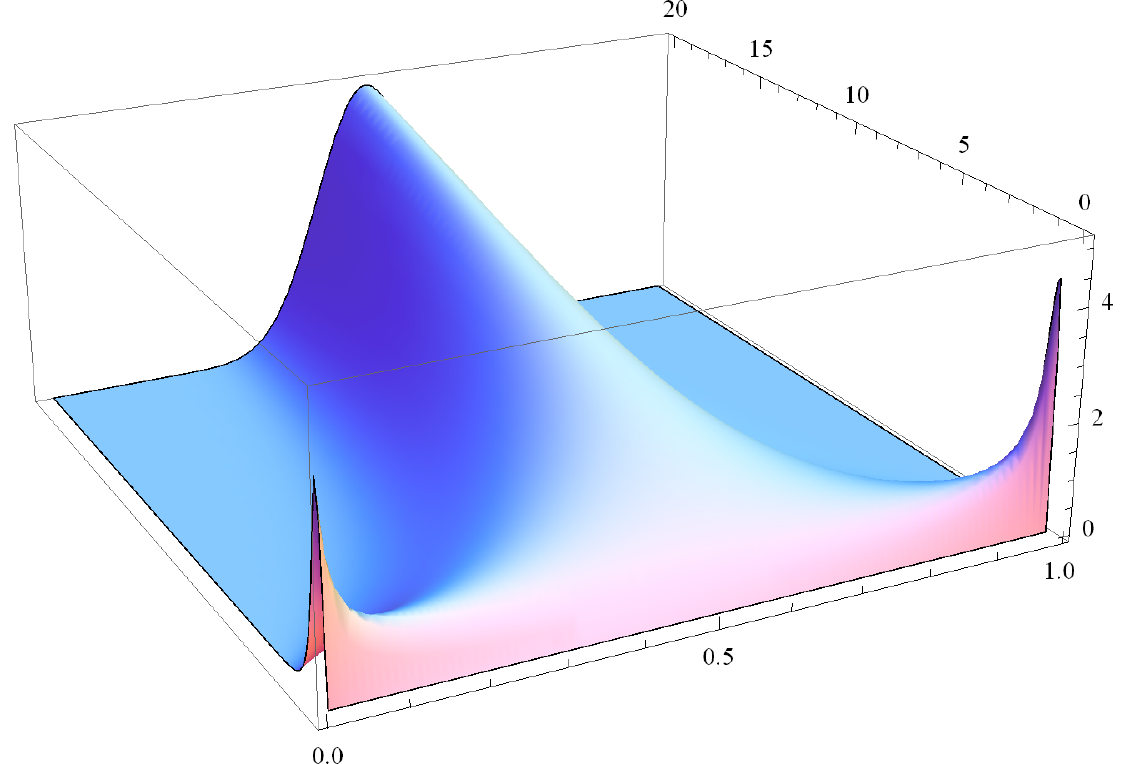}}}
\put(80,55){\large\bf Central mean $P_0(a,\nu,\tau)$}
\put(235,140){\large\bf $\tau$}
\put(130,-7){\large\bf Unfitness $a$}
\end{picture}
\end{center} \caption{{\em The log-gamma family of probability
densities $P_0(a,\nu,\tau)$ from (\ref{loggam}) as a surface for the case of
central mean $E_0(a) \, = \frac{1}{2}.$
This symmetric surface coincides with the uniform density $1$ at $\tau=1,$ and tends to the
delta function as $\tau\rightarrow\infty$}}
\label{LGF}
\end{figure}
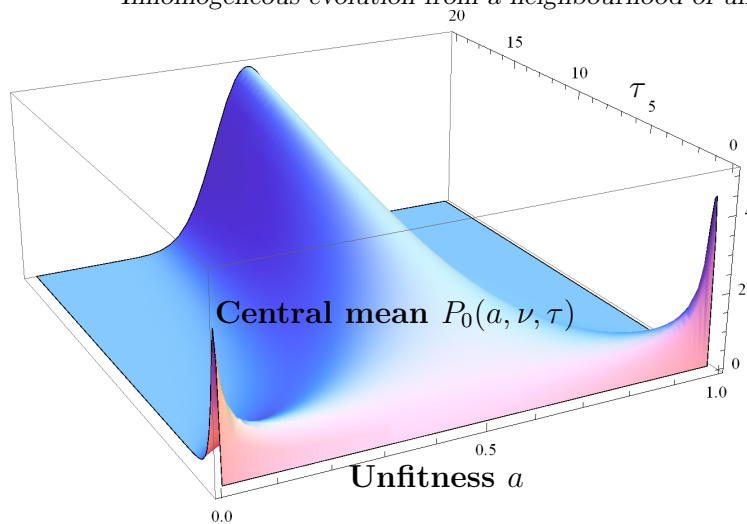
In the example we elaborate, the inhomogeneous population $N$
is classified by a smooth family of probability density functions
$\{P_t, t\geq 0\}$ with random variable $0\leq a \leq 1,$  having mean $E_t(a)$ and variance
$\sigma_t^2(a)=E_t(a^2)-(E_t(a))^2.$
Here $a$ represents an unfitness that controls the decline in the frequency of the $a$-cohort,
so we could view fitness as the variable $1-a.$

Let $l_t(a)$ represent the frequency at the $a$-cohort, then we have
\begin{eqnarray}
   N(t)&=& \int_0^\infty l_t(a) \, da  \ \ \ {\rm and} \ \ P_t(a) =\frac{l_t(a)}{N(t)}\\
                         \frac{dl_t(a)}{dt}&=&  -a l_t(a)  \ \ \ {\rm so} \ \ l_t(a) = l_0(a) e^{-at}
                         \end{eqnarray}
Karev \cite{Karev03}  obtained general solutions for
these equations giving us
\begin{eqnarray}
                         N(t) &=& N(0)L_0(t) \ \ {\rm where} \
                         L_0(t)= \int_0^\infty P_0(a) e^{-at} \, da \label{L0}\\
                         \frac{dN}{dt}&=& -E_t(a) \, N \ \ {\rm where} \
                         E_t(a)= \int_0^\infty a \, P_t(a) \, da  = -\frac{d\log L_0}{dt} \label{Eta}\\
                          \frac{dE_t(a)}{dt}&=& -\sigma_t^2(a) = (E_t(a))^2 - E_t(a^2) \label{Vta}\\
                         P_t(a) &=& e^{-at}\frac{P_0(a)}{L_0(t)} \ \ {\rm and} \ l_t(a) = e^{-at}L_0(t) \label{Pt}\\
                        \frac{dP_t(a)}{dt}&=&  P_t(a) (E_t(a)-a). \label{Prate}
                       \end{eqnarray}
Here $L_0(t)$ is the Laplace transform of
the initial probability density function $P_0(a)$ (which is of course zero outside $[0,1]$)
and so conversely $P_0(a)$ is the
inverse Laplace transform of the population (monotonic) decay solution $\frac{N(t)}{N(0)}.$ See
Feller~\cite{Feller} for more discussion of the existence and uniqueness properties of the
correspondence between probability densities and their Laplace transforms.
We see from (\ref{Prate}) that when the unfitness $a$ exceeds its mean $E_t(a)$ then the population
density of that cohort declines and conversely the densities of cohorts with $a<E_t(a)$ tend to grow.
The Shannon entropy at time $t$ is
\begin{equation}\label{Sdef}
    S_t =  - E_t\left(\log P_t(a)\right) =
    - E_t\left(\log \frac{P_0(a) e^{-at}}{L_0(t)} \right)
\end{equation}
which reduces to
\begin{equation}\label{S}
    S_t = S_0 + \log L_0(t) + E_t(a) \, t .
\end{equation}
By using $\frac{dE_t}{dt} = -\sigma_t^2(a),$ the decay rate is then
\begin{equation}\label{Sdecay}
    \frac{dS_t}{dt} = -\sigma^2(t) \, t.
\end{equation}
This shows how the variance controls the entropy change during
quite general inhomogeneous population processes, as we saw in~\cite{stochrateproc}.

Karev gave the particular solutions for the cases of initial densities that were
Poisson, gamma or uniform. In~\cite{stochrateproc} we studied the case of a bivariate
gamma density    for an epidemic situation. Here we shall study a family of log-gamma
densities that determine a neighbourhood of the uniform distribution~\cite{InfoGeom},
so recovering the solution of Karev for the uniform distribution as a special case ($\nu=\tau=1$).
The log-gamma
family contains also close approximations to Gaussians of arbitrarily small variance
truncated to $[0,1]$ (for $\tau>1$) as well as
densities having outlier cohorts of extreme values that are enhanced ($\tau<1$), Figure~\ref{LGF}.
\begin{figure}
\begin{picture}(300,130)(0,0)
\put(60,0){\resizebox{8cm}{!}{\includegraphics{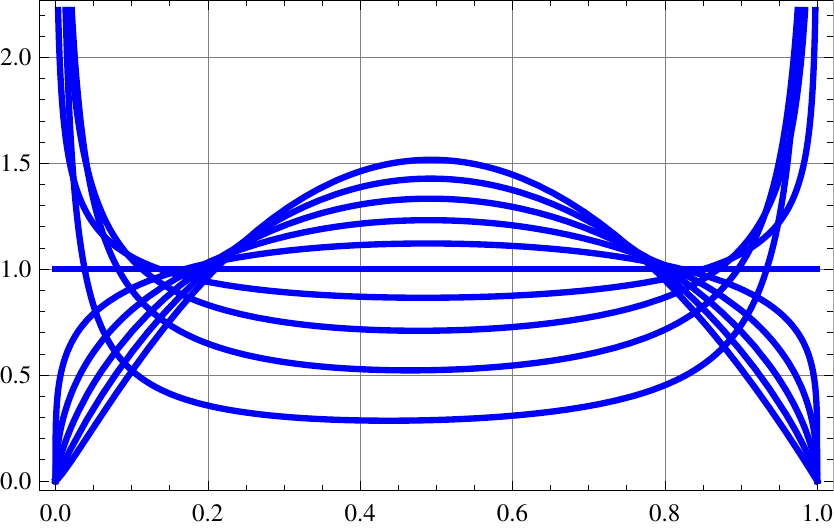}}}
\put(32,128){\large\bf $P_0(a)$}
\put(85,120){\large \textbf{$\tau <1$}}
\put(170,33){\large \textbf{$\tau <1$}}
\put(170,105){\large \textbf{$\tau >1$}}
\put(150,-5){Unfitness $a$}
\end{picture}
\caption{{\em Log-gamma probability density functions $P_0(a)$
from (\ref{loggam}) for $a\in [0,1],$ with central
mean $E_0(a)=\frac{1}{2},$ and $\tau$ from $0.2$ to $2$ in steps of $0.2.$
Note that the parameter $\tau$ controls the shape of the graph and for
$\tau=1$ we have $P_0(a)=1.$}}
\label{sections}
\end{figure}
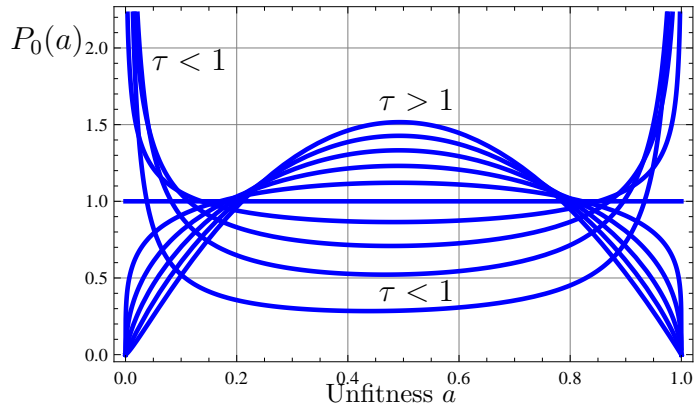

\begin{figure}
\begin{center}
\begin{picture}(300,130)(0,0)
\put(-25,0){\resizebox{6cm}{!}{\includegraphics{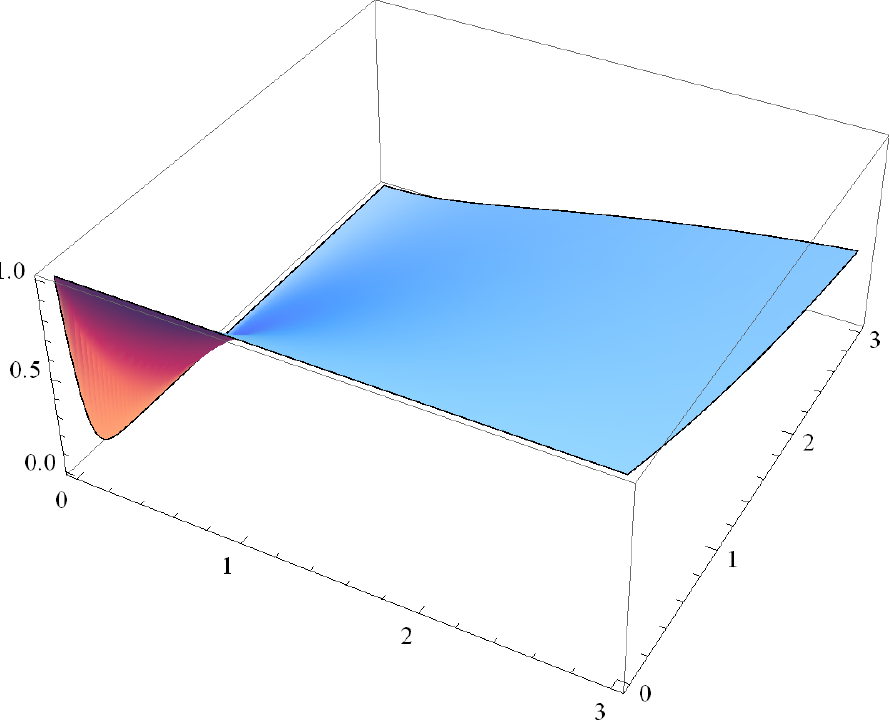}}}
\put(150,-10){\resizebox{6cm}{!}{\includegraphics{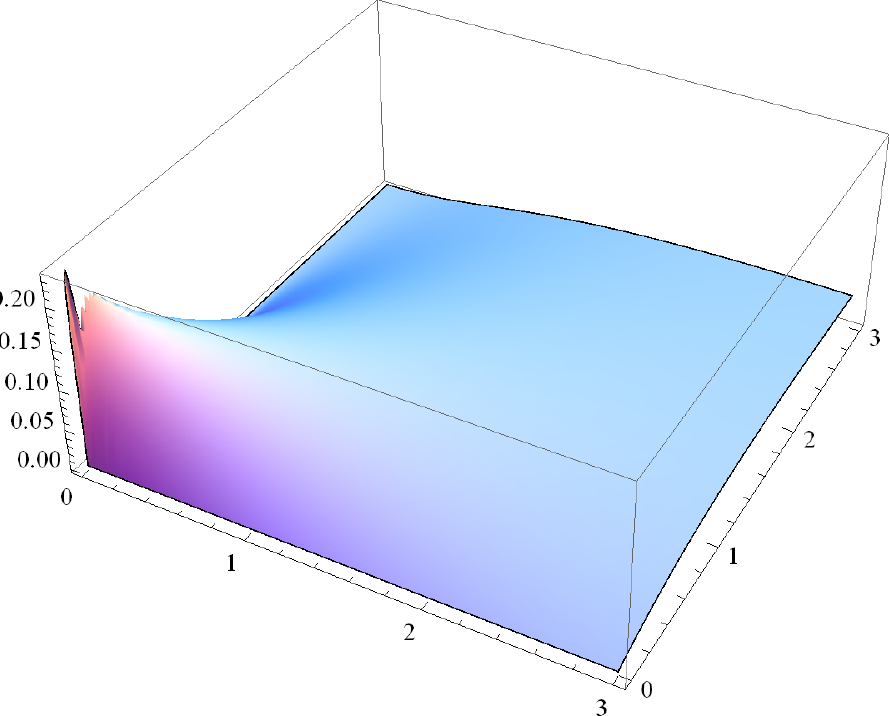}}}
\put(0,110){{ Log-gamma mean}}
\put(150,110){{ Log-gamma variance}}
\put(120,30){{\large $\tau$}}
\put(65,6){\large $\nu$}
\put(310,30){{\large $\tau$}}
\put(250,-10){\large $\nu$}
\end{picture}
\end{center}
\caption{{\em Mean and variance for the
log-gamma family.}}
\label{LogGamMVar}
\end{figure}
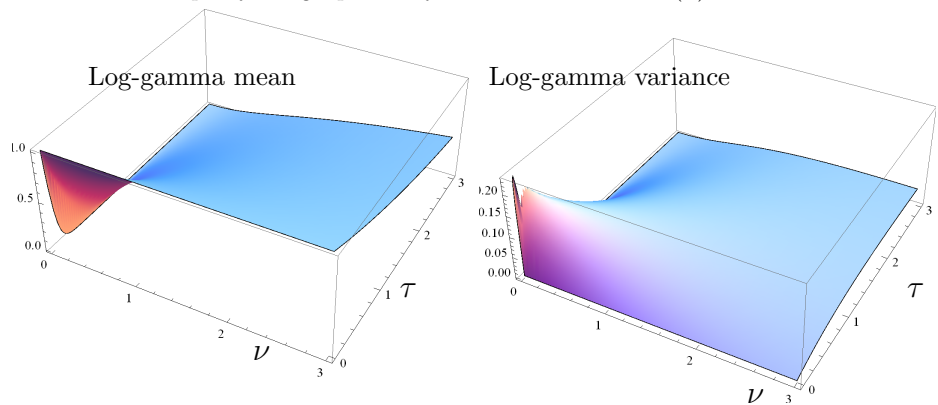

\section{Evolution of inhomogeneity from an initial log-gamma probability density}
\begin{figure}
\begin{center}
\begin{picture}(300,180)(0,0)
\put(0,-10){\resizebox{10 cm}{!}{\includegraphics{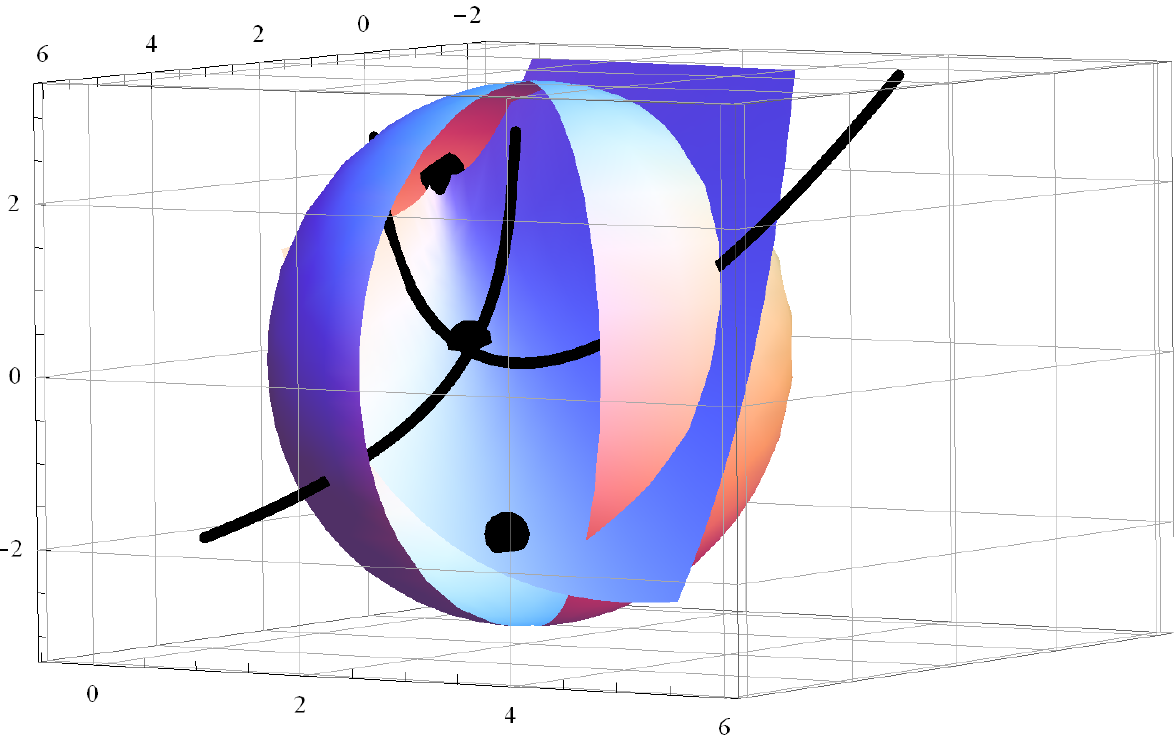}}}
\put(90,-10){\large $\nu$}
\put(50,160){\large $\tau$}
\put(213,140){\large $\tau=1$}
\put(18,40){\large $\nu=1$}
\end{picture}
\end{center} \caption{{\em An affine immersion in $\mathbb{R}^3$ of the 2-manifold of log-gamma
probability densities. The black curves
in the immersion represent the distributions with $\nu=1$ and $\tau=1$ and centred on
their intersection is a spherical neighbourhood of the uniform distribution. The other two points
are for the two cases $(\nu=0.1,\tau=0.289)$ and $(\nu=2.75,\tau=2.24)$
 cf.  Figures~\ref{EtaDec},\ref{L0Dec},\ref{VtaDec}}}
\label{SphNhd}
\end{figure}
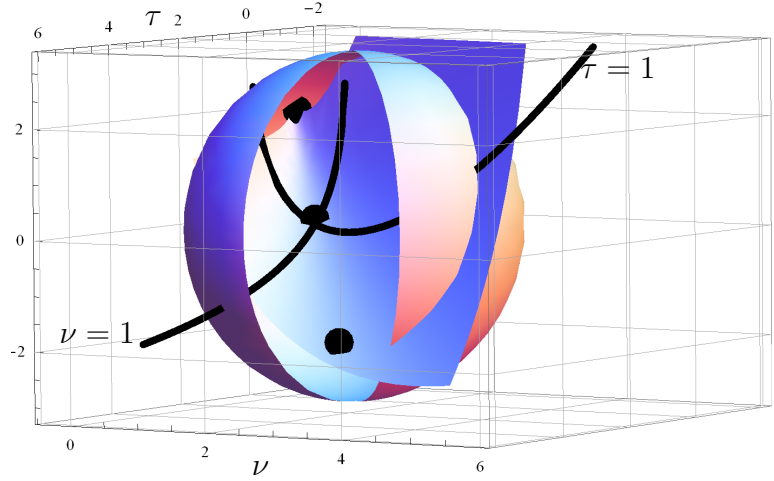

\begin{figure}
\begin{picture}(300,90)(0,0)
\put(175,0){\resizebox{6cm}{!}{\includegraphics{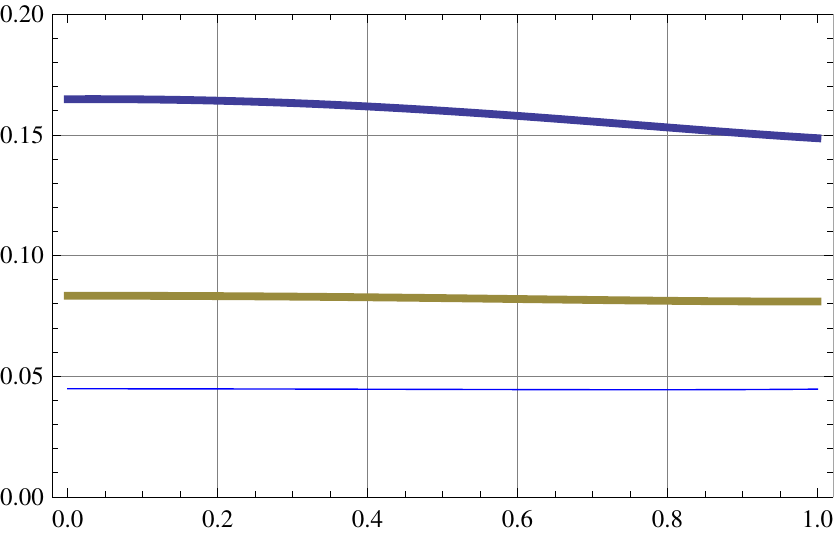}}}
\put(0,0){\resizebox{6cm}{!}{\includegraphics{P0aExs}}}
\put(15,93){\large\bf $P_0(a)$}
\put(191,93){$\sigma^2_t(a)$}
\put(235,-7){\large\bf Time $t$}
\put(65,-7){\large\bf Unfitness $a$}
\end{picture}
\caption{{\em Initial log-gamma densities $P_0(a)$ shown in left panel with central
mean $E_0(a)=\frac{1}{2}$ for the cases $\tau=0.289,$  $\nu=0.1$ (lower graph),
 $\tau=2.24,$  $\nu=2.75$ (upper graph), and the uniform density  $P_0(a)=1$
 for  $\tau=\nu=1.$
 The right panel shows the decline with time of the variance $\sigma^2_t(a)$
from (\ref{Vta}) for these initial densities.}}
\label{VtaDec}
\end{figure}

We studied in~\cite{InfoGeom} the smooth family of log-gamma distributions and their
information geometry. This family has probability density function
\begin{equation}\label{loggam}
    P_0(a,\nu,\tau) =   \frac{a^{\nu -1} \nu ^{\tau } \left|\log
   \left(\frac{1}{a}\right)\right|^{\tau -1}}{\Gamma (\tau )}
\end{equation}
for random variable $a\in [0,1]$ and parameters $\nu,\tau >0.$ The mean and variance,
Figure~\ref{LogGamMVar}, are given by
\begin{eqnarray}
E_0(a)&=&\left( \frac{\nu }{1  + \nu } \right)^\tau \\
\sigma_0^2(a) &=&  \left(\frac{\nu }{\nu  + 2 }\right)^
       \tau- \left(\frac{\nu }{1  + \nu }\right)^{2 \tau } .
\end{eqnarray}
The locus in this family of those with central mean $E_0(a)=\frac{1}{2}$ satisfies
\begin{equation}\label{centmean}
\nu (2^{\frac{1}{\tau }}-1)=1
\end{equation}
and some are shown in Figure~\ref{LGF}
The uniform density    is the special case with $\tau= \nu=1$ and this can be seen
in Figure~\ref{sections}, with sections through the surface of Figure~\ref{LGF}.
The log-gamma family yields a smooth Riemannian 2-manifold with coordinates
$(\nu,\tau)\in \mathbb{R}^+ \times \mathbb{R}^+$ and metric tensor given by
\begin{equation}\label{loggammetric}
\left[g_{ij}\right](\nu,\tau)=\left[ \begin{array}{cc}
        \frac{\tau}{{\nu}^2}  &    -\frac{1}{\nu}  \\
        -\frac{1}{\nu}   &   \frac{d^2}{d\tau^2}\log(\Gamma)
\end{array} \right].
\end{equation}
In fact, this manifold is an isometric diffeomorph of the 2-manifold of gamma densities,
with random variable $x=-\log a$ via natural coordinates $(\nu,\tau)$ (cf.~\cite{AN})
for which $\tau=1$ corresponds to the subfamily of exponential densities.
Through this smooth diffeomorphism we can therefore represent the manifold of log-gamma densities
 as a natural affine immersion (cf.~\cite{InfoGeom})
of $\mathbb{R}^+ \times \mathbb{R}^+$ in $\mathbb{R}^3$
\begin{equation}\label{affim}
    (\nu,\tau)\mapsto \{\nu,\tau,\log\Gamma(\tau) -\tau\log \nu\}.
\end{equation}
This is illustrated in Figure~\ref{SphNhd} which shows also a spherical neighbourhood in $\mathbb{R}^3$ centred
on the point at the uniform density, $\nu=1=\tau;$  the other two points
are for the two cases $(\nu=0.1,\tau=0.289)$ and $(\nu=2.75,\tau=2.24)$ which represent densities
also having mean value $\frac{1}{2}$ and used in the sequel as initial unfitness distributions,
cf.  Figures~\ref{EtaDec},\ref{L0Dec},\ref{VtaDec}.

The log-gamma entropy (\ref{Sdef}) is given by
\begin{eqnarray}
   S_{LG}(\nu,\tau)&=&\nu ^{\tau } (\nu +1)^{-\tau -1}\left(  A(\nu,\tau) +B(\nu,\tau)    \right)\label{LGent} \\
  {\rm with} \ A(\nu,\tau) &=& \tau  (\nu +(\nu +1) \log (\nu +1)-1)-(\nu +1) (\tau -1) \psi(\tau )\nonumber \\
   {\rm and} \  B(\nu,\tau) &=& (\nu +1) \log \left(\frac{\nu ^{-\tau } \Gamma (\tau )}{\nu +1}\right),
   \ \psi(\tau )= \frac{d\log{\Gamma(\tau)}}{d\tau} .\nonumber
\end{eqnarray}
\begin{figure}
\begin{center}
\begin{picture}(300,135)(0,0)
\put(-40,-5){\resizebox{7cm}{!}{\includegraphics{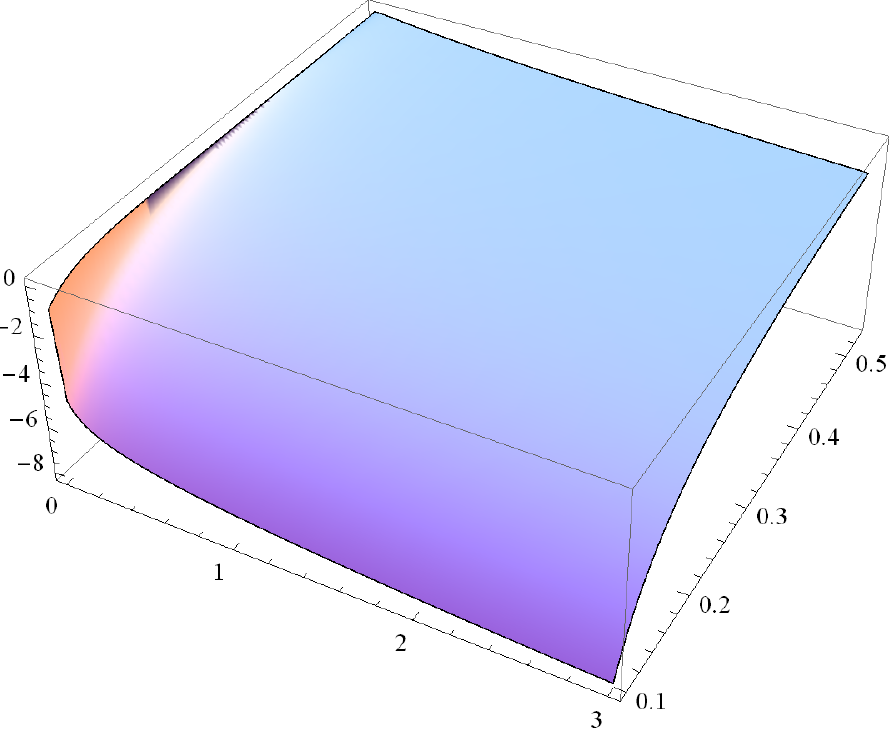}}}
\put(165,-15){\resizebox{6cm}{!}{\includegraphics{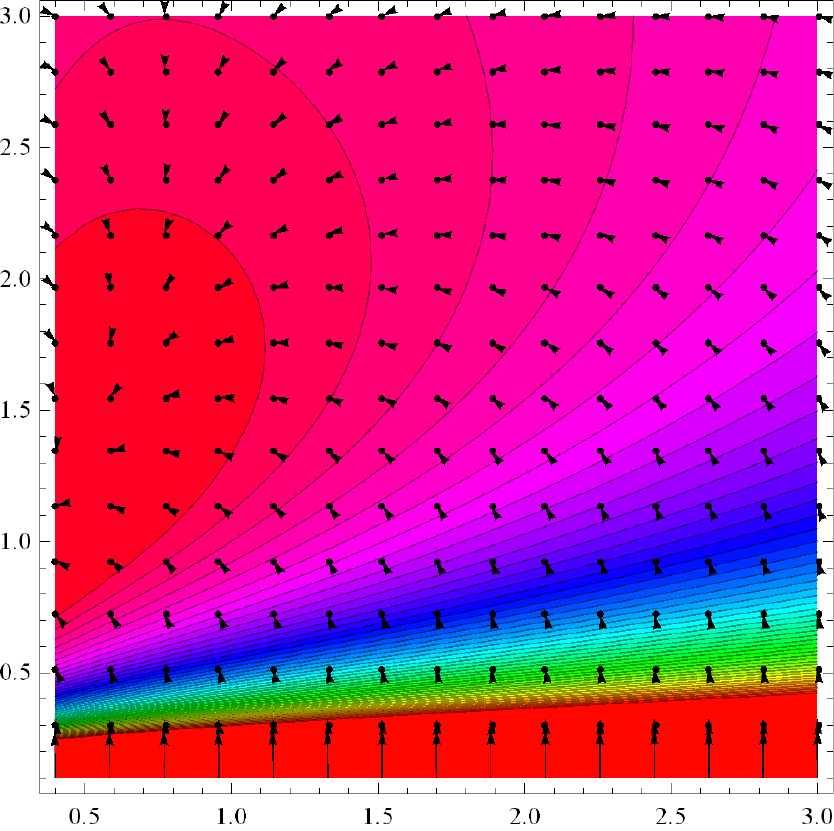}}}
\put(20,115){{\large Log-gamma entropy}}
\put(140,45){{\large $\tau$}}
\put(55,5){\large $\nu$}
\put(163,135){{\large $\tau$}}
\put(220,-15){\large $\nu$}
\end{picture}
\end{center}
\caption{{\em Shannon entropy function for the
log-gamma family as a surface (left) and
as a contour plot with entropy gradient flow (right).}}
\label{LogGamEntSurfCont}
\end{figure}
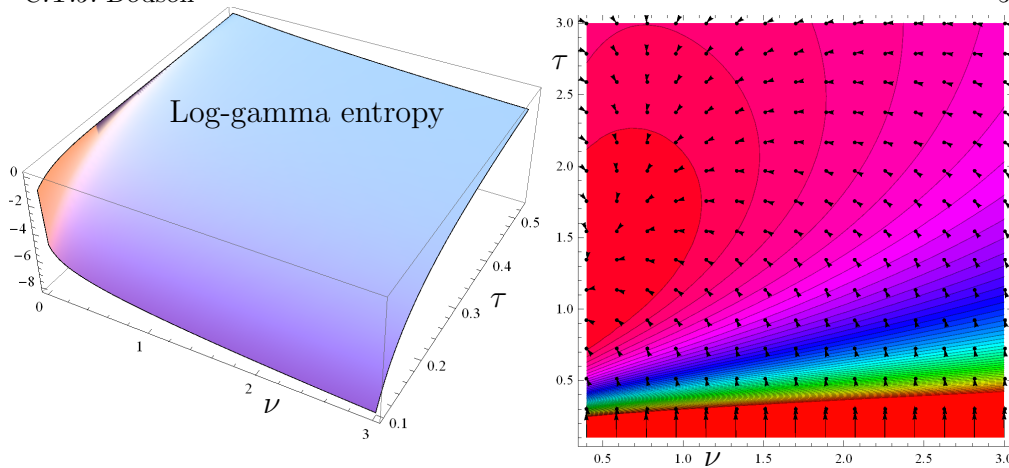
Figure~\ref{LogGamEntSurfCont}
shows the graph as a surface and as a contour plot with gradient vector
field indicated; the main curvature occurs near the origin.

The Laplace transform integral (\ref{L0}) for the log-gamma density with general $(\nu,\tau)$
seems intractable so we used a series
development up to sixth order for the term $ e^{-at},$ bearing in mind that $a\in [0,1].$ Thus, up to sixth
order in $t\geq 0$ we obtain a good approximation to $L_0(t),$ accurate to about $0.1\%$ up to $t=1.$
The corresponding expressions for $L_0(t), \ P_t(a), \ E_t(a), \ \sigma_t^2(a)$ and hence $N(t)$ are
known but somewhat cumbersome to present here so we present some graphics for illustration.
The decline of the mean $E_t(a)$ as the density    $P_t(a)$ develops
from three initial log-gamma densities $P_0(a)$ with central mean, $E_0(a)=\frac{1}{2}$
is shown in Figure~\ref{EtaDec}. The more rapid decline occurs for the
initial density    with $\tau =0.289$ and the slower decline occurs for
the case having $\tau = 2.24$ with the uniform case $\tau=\nu=1$ in between,
cf. also Figure~\ref{SphNhd}.
The corresponding decline  of $N(t)/N(0)=L_0(t)$ is shown in Figure~\ref{L0Dec} and
the variance is shown in Figure~\ref{VtaDec}.

We do have an analytic solution for the special 1-parameter family of log-gamma densities
with $\tau=1$ (the \lq log-exponential'  densities), from (\ref{L0}) and (\ref{Pt}):
\begin{eqnarray}
    P_t(a,\nu,1) &=& \frac{e^{-a t} a^{\nu -1} t^{\nu }}{\Gamma (\nu )-\Gamma (\nu ,t)}\label{Ptnu} \\
    E_t(a,\nu,1) &=& \frac{\Gamma (\nu +1)-\Gamma (\nu +1,t)}{t(\Gamma (\nu )-\Gamma (\nu
   ,t))} \label{Etnu}\\
  \frac{N(t)}{N(0)} &=& L_0(t)=\nu  t^{-\nu } (\Gamma (\nu )-\Gamma (\nu ,t))\label{Ntnu}
\end{eqnarray}
where $\Gamma (\nu ,t)$ is the incomplete gamma function. $E_t(a,\nu,1),$
declines for all choices of parameter $\nu$ in the initial density,
which includes of course also the uniform density. Further analytic solutions arise
for other initial log-gamma densities with positive integer $\tau$
(the \lq log-Pearson Type III' distributions)
cf. Figure~\ref{sections},
as generalized hypergeomentric functions;  for example,
\begin{eqnarray}
 \tau=2 & \Rightarrow &  \frac{N(t)}{N(0)} = L_0(t)=  \, _2F_2(\nu ,\nu ;\nu +1,\nu +1;-t)\label{t2}\\
 \tau=3 & \Rightarrow &  \frac{N(t)}{N(0)} = L_0(t)=  \, _3F_3(\nu ,\nu ,\nu ;\nu +1,\nu +1,\nu +1;-t)\label{t3}.
\end{eqnarray}
\begin{figure}
\begin{picture}(300,110)(0,0)
\put(175,0){\resizebox{6cm}{!}{\includegraphics{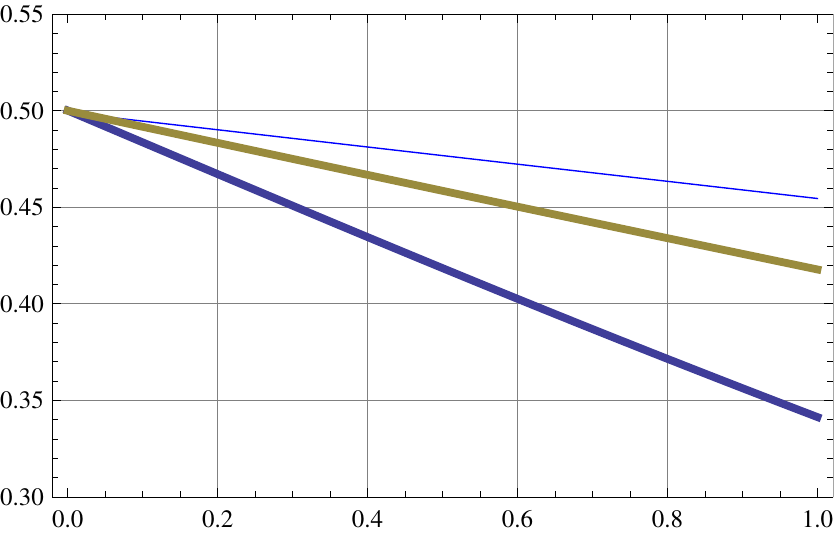}}}
\put(0,0){\resizebox{6cm}{!}{\includegraphics{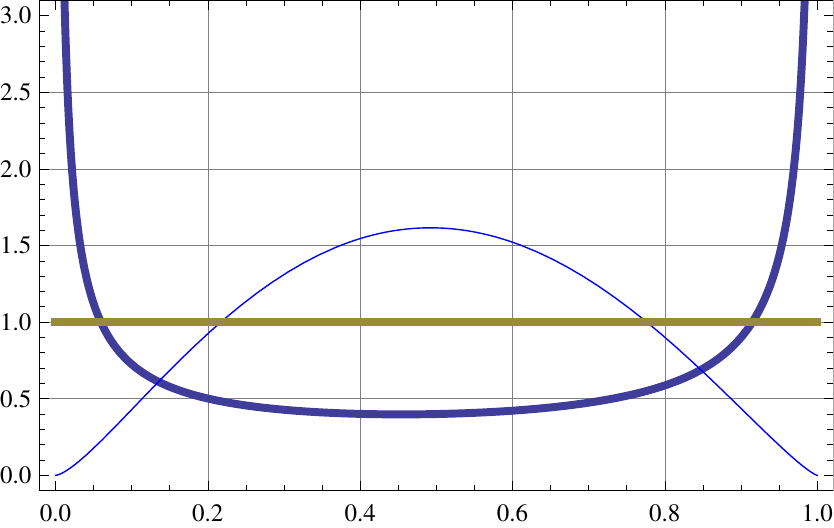}}}
\put(15,93){\large\bf $P_0(a)$}
\put(191,90){\large\bf $E_t(a)$}
\put(235,-7){\large\bf Time $t$}
\put(65,-7){\large\bf Unfitness $a$}
\end{picture}
\caption{{\em Initial log-gamma densities $P_0(a)$ shown in left panel with central
mean $E_0(a)=\frac{1}{2}$ for the cases $\tau=0.289,$  $\nu=0.1$ (lower graph),
 $\tau=2.24,$  $\nu=2.75$ (upper graph), and the uniform density      $P_0(a)=1$
 for  $\tau=\nu=1.$ The right panel shows the decline with time of the mean $E_t(a)$
from (\ref{Eta}) for these initial densities.}}
\label{EtaDec}
\end{figure}
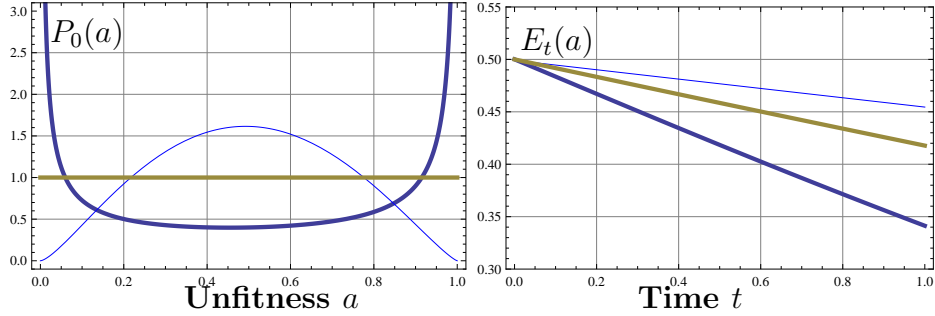
\begin{figure}
\begin{picture}(300,110)(0,0)
\put(175,0){\resizebox{6cm}{!}{\includegraphics{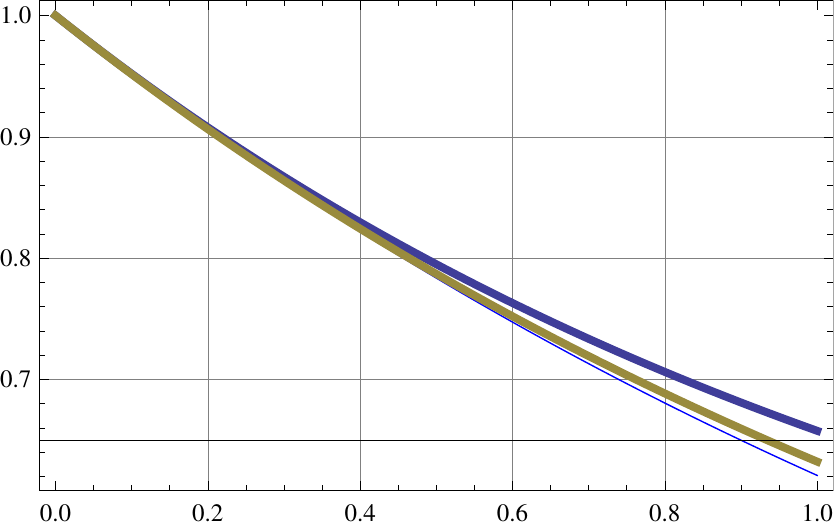}}}
\put(0,0){\resizebox{6cm}{!}{\includegraphics{P0aExs}}}
\put(15,93){\large\bf $P_0(a)$}
\put(210,93){\large\bf $N(t)/N(0)=L_0(t)$}
\put(235,-7){\large\bf Time $t$}
\put(65,-7){\large\bf Unfitness $a$}
\end{picture}
\caption{{\em Initial log-gamma densities $P_0(a)$ shown in left panel with central
mean $E_0(a)=\frac{1}{2}$ for the cases $\tau=0.289,$  $\nu=0.1$ (lower graph),
 $\tau=2.24,$  $\nu=2.75$ (upper graph), and the uniform density  $P_0(a)=1$
 for  $\tau=\nu=1.$
 The right panel shows the fractional decline with time of the population $N(t)/N(0)=L_0(t)$
from (\ref{L0}) for these initial densities.}}
\label{L0Dec}
\end{figure}

\begin{figure}
\begin{picture}(300,120)(0,0)
\put(185,0){\resizebox{6.5cm}{!}{\includegraphics{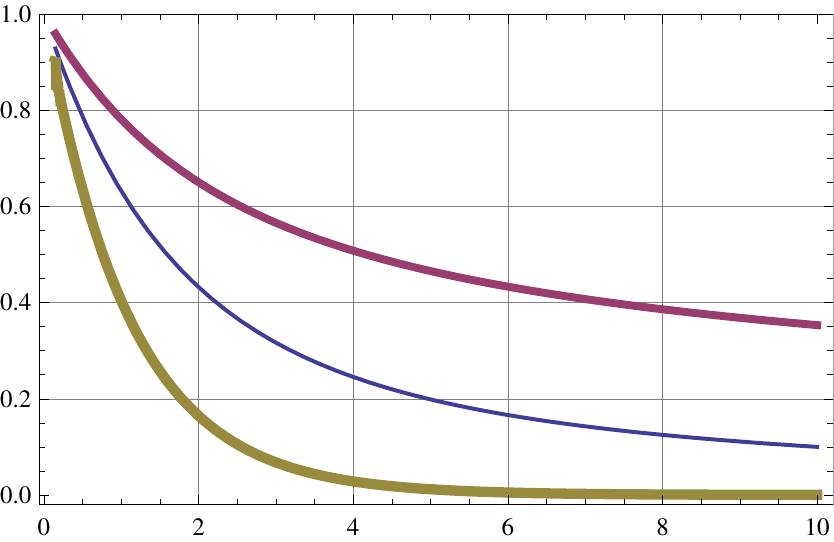}}}
\put(0,0){\resizebox{6.4cm}{!}{\includegraphics{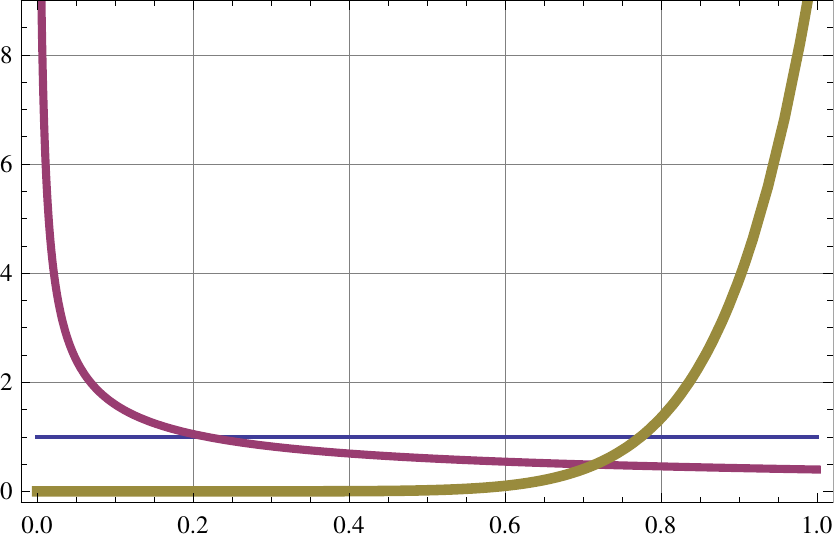}}}
\put(15,108){\large\bf $P_0(a,\nu,1)$}
\put(12,74){$\nu=0.4$}
\put(140,93){$\nu=10$}
\put(79,23){$\nu=1$}
\put(231,103){\large\bf $N(t)/N(0)=L_0(t)$}
\put(248,42){$\nu=1$}
\put(197,23){$\nu=10$}
\put(232,78){$\nu=0.4$}
\put(235,-7){\large\bf Time $t$}
\put(65,-7){\large\bf Unfitness $a$}
\end{picture}
\caption{{\em Initial log-gamma densities $P_0(a,\nu,1)$ from (\ref{Ptnu}) are shown in the
left panel  for the uniform density
 $\tau=\nu=1,$ and also for $\tau=1$ with $\nu=0.4$ and $10;$ Figure~\ref{Ptnu1} shows the evolution
 of the case with $\nu=10.$
 The right panel shows the corresponding fractional decline with time of the population $N(t)/N(0)$
from (\ref{Ntnu}) for these initial densities. Figure~\ref{Ptnu1} shows the evolution from $P_0(a,10,1).$}}
\label{NTau1Dec}
\end{figure}
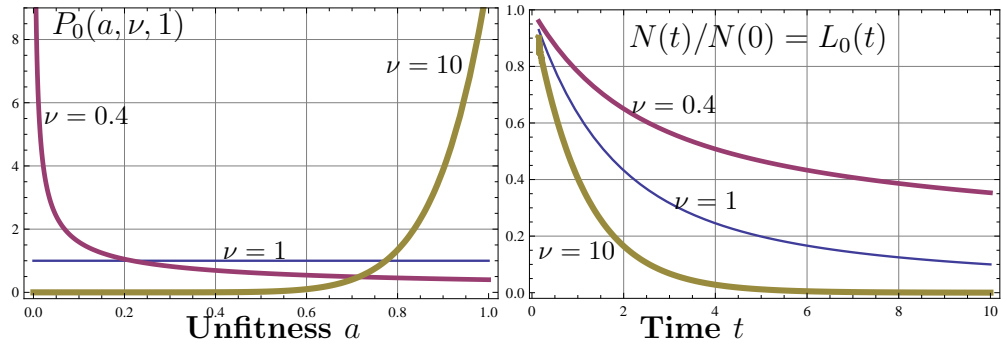
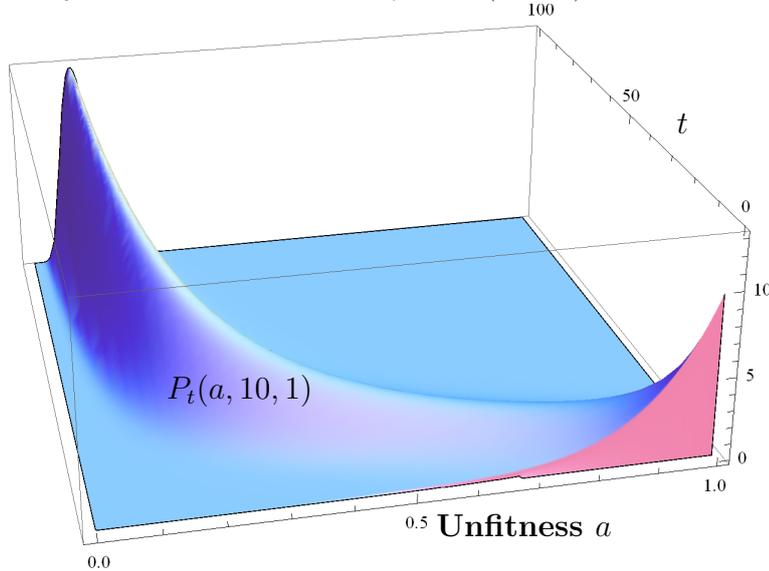
\begin{figure}
\begin{center}
\begin{picture}(300,165)(0,0)
\put(0,-19){\resizebox{10 cm}{!}{\includegraphics{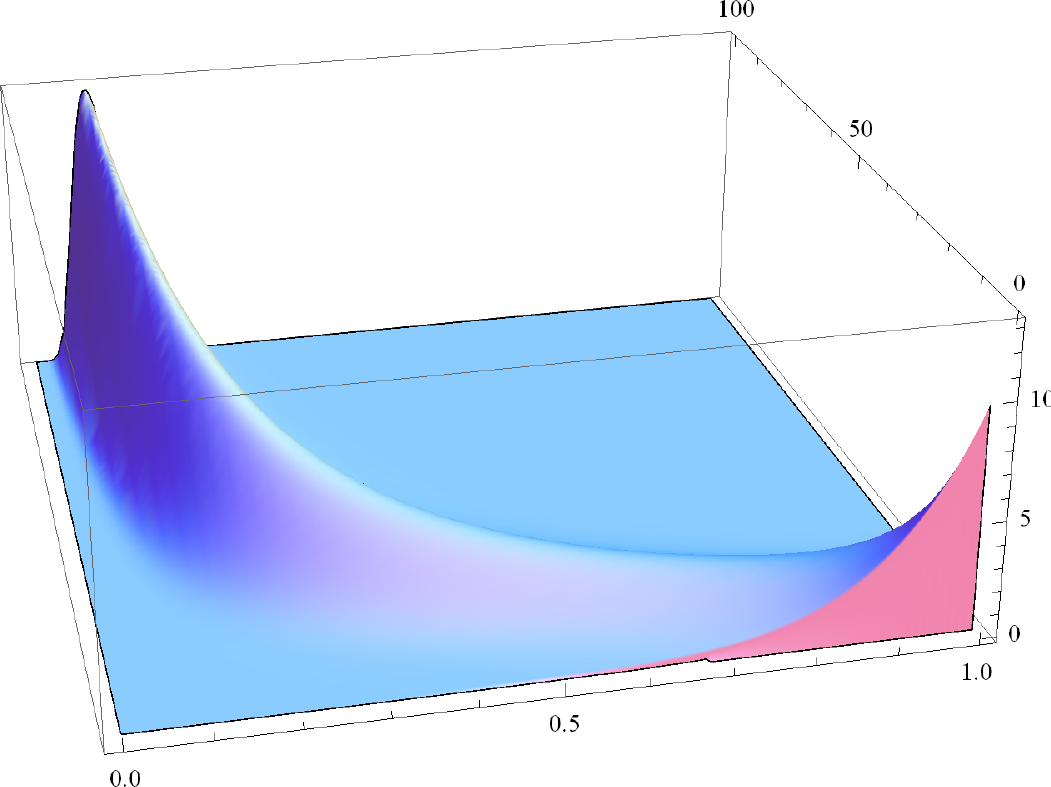}}}
\put(59,45){\large\bf $P_t(a,10,1)$}
\put(250,145){\large\bf $t$}
\put(160,-7){\large\bf Unfitness $a$}
\end{picture}
\end{center} \caption{{\em The evolution of probability
density $P_t(a,10,1)$ from (\ref{Ptnu}).}}
\label{Ptnu1}
\end{figure}
\begin{figure}
\begin{picture}(300,120)(0,0)
\put(185,0){\resizebox{6.5cm}{!}{\includegraphics{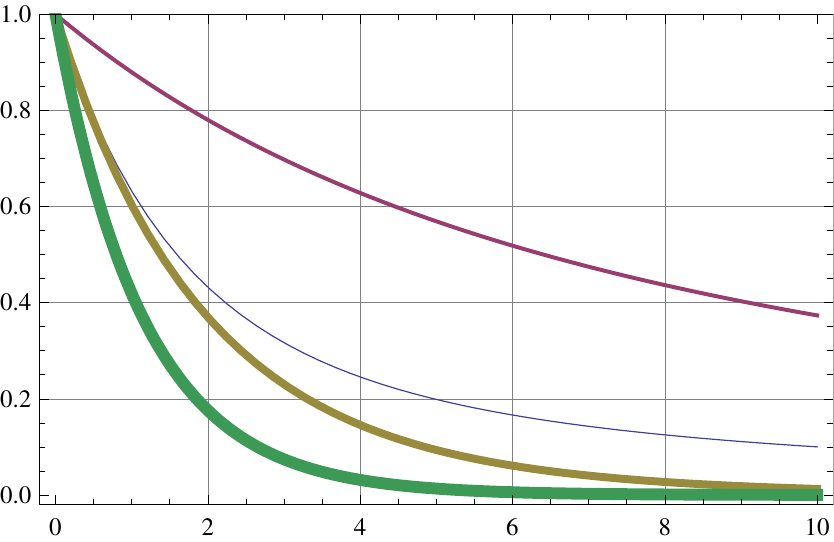}}}
\put(0,0){\resizebox{6.4cm}{!}{\includegraphics{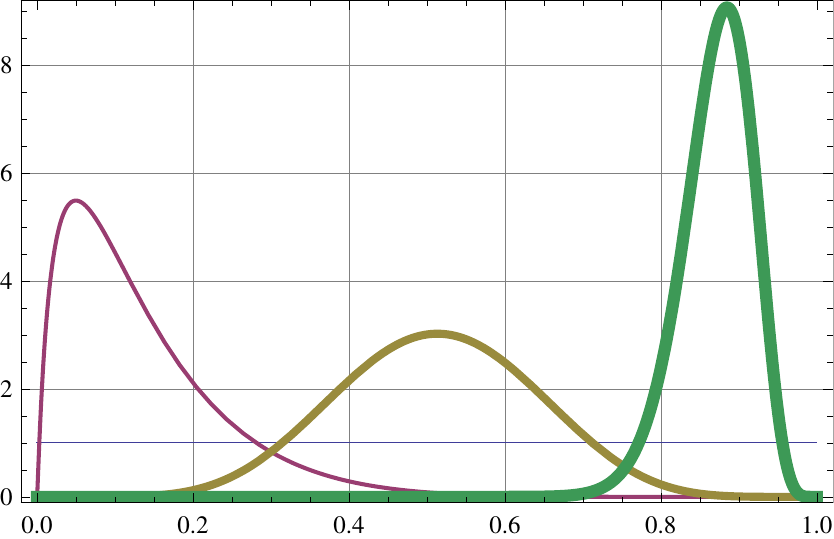}}}
\put(15,108){\large\bf $P_0(a,\nu,7)$}
\put(18,73){$\nu=3$}
\put(105,43){$\nu=10$}
\put(125,93){$\nu=50$}
\put(79,23){$\nu=1$}
\put(231,103){\large\bf $N(t)/N(0)=L_0(t)$}
\put(248,42){$\nu=1$}
\put(248,20){$\nu=10$}
\put(198,23){$\nu=50$}
\put(290,67){$\nu=3$}
\put(235,-7){\large\bf Time $t$}
\put(65,-7){\large\bf Unfitness $a$}
\end{picture}
\caption{{\em Initial log-gamma densities $P_0(a,\nu,7)$ are shown in the left panel
for the uniform density
 $\tau=\nu=1,$ and also for $\tau=7$ with $\nu=3, \ 10, \ 50$ in graphs with increasing thickness.
 The right panel shows the corresponding fractional decline with time of the population $N(t)/N(0)$
from (\ref{NTau7}) for these initial densities.}}
\label{NTau7Dec}
\end{figure}
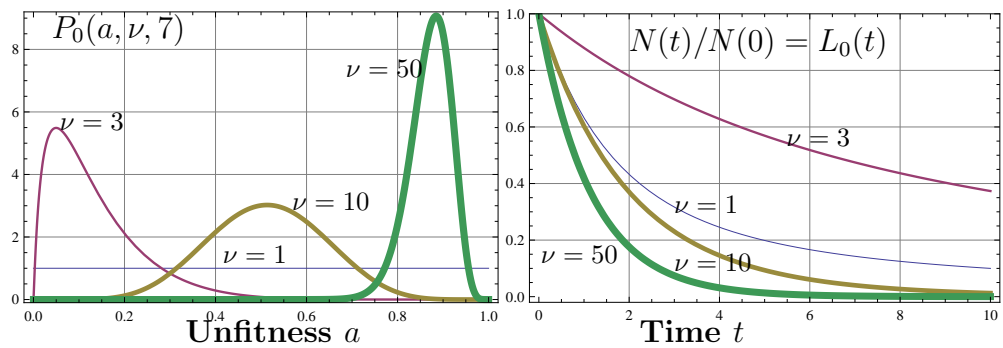
\newpage\section{Discussion}
These new results show
how evolution of an inhomogeneous rate process reacts to perturbations of the uniform density
using the smooth two parameter ($\nu,\tau$) family of log-gamma probability
density functions~\cite{InfoGeom} to represent unfitness measured
by parameter $a\in [0,1].$ We have illustrated the evolution from three initial states having
central mean $E_0(a)=\frac{1}{2}$ but different variances (cf. Figures~\ref{sections} and \ref{SphNhd}):
\begin{description}
\item [initial uniformity] ($\tau=1$ with variance $\sigma_0^2(a)=\frac{1}{12}\approx 0.083$)
\item [initial higher variance] ($\tau=0.289$ with  $\sigma_0^2(a)=0.164$)
\item [initial lower variance] ($\tau=2.24$ with  $\sigma_0^2(a)=0.045$).
\end{description}
The graphs of these three starting log-gamma densities in the left hand sides of
Figures~\ref{EtaDec},\ref{L0Dec},\ref{VtaDec}
show widely differing shapes; the higher variance arises from peaks at the extremities of unfitness
and the lower variance case has a symmetric bell-like curve about the mean.
The evolution of features from the initial uniform density
from Karev~\cite{Karev03} is---cf. equation (\ref{Pt}) above,
\begin{equation}\label{K0}
    P_0(a)=1 \Rightarrow P_t(a)=\frac{t e^{-a t}}{1-e^{-t}}, \ \ E_t(a)=\frac{1}{t}+\frac{1}{1-e^t}, \ \
    \frac{N(t)}{N(0)}= \frac{1-e^{-t}}{t}.
\end{equation}
To this we add our analytic results (\ref{Ptnu}),  (\ref{Etnu}) for the initial log-gamma cases
with $\tau=1,$ which agrees with (\ref{K0}) at $\nu=1$ and the generalized hypergeometric solutions
(\ref{t2}), (\ref{t3}) along with corresponding versions for integer $\tau>3.$
Our series approximation agrees with the analytic results for the early development.
Figure~\ref{EtaDec} shows that in each of our cases the
evolution of mean unfitness $E_t(a)$ does indeed follow the expected equation (\ref{Vta}),
which expresses Fisher's law of natural selection~\cite{Fisher}.
Also, in Figure~\ref{VtaDec} the very slow early evolution of its variance $\sigma_t^2(a)$
reflects the corresponding lack of curvature in the mean.
In fact, though the decline of the mean fitness does change with the variation in starting conditions,
the net population evolution $N(t)$ in Figure~\ref{L0Dec} seems surprisingly stable under quite
considerable alterations in the shape and variance of the initial density  of inhomogeneity
in fitness. The rate process evidently has a strong smoothing effect when we begin from a central
mean, even though the variances differ widely.

We can however obtain considerable changes in the evolution if we depart from an initial
central mean. We illustrate this using our analytic solutions to allow evolution
over longer periods. For example, the initial log-gamma density  for $\tau=1,$
equations (\ref{Ptnu}),(\ref{Etnu}),(\ref{Ntnu}), gives the examples shown in Figure~\ref{NTau1Dec}.
Here we see the effect of $\nu$ which acts as a location parameter for the initial density.
Figure~\ref{Ptnu1} shows the evolution of the initial case with $\nu=10$ into the probability
densities $P_t(a,10,1)$ from (\ref{Ptnu}).

Also, the initial unimodular log-gamma  probability
densities for $\tau=7,$  $P_t(a,\nu,7),$ give this generalized hypergeometric solution
\begin{equation}\label{NTau7}
   \frac{N(t)}{N(0)} = L_0(t) =\, _7F_7(\nu ,\nu ,\nu ,\nu ,\nu ,\nu ,\nu ;\nu +1,\nu +1,\nu +1,\nu +1,\nu +1,\nu +1,\nu +1;-t)
\end{equation}
and again $\nu$ serves as a location parameter.
Figure~\ref{NTau7Dec} illustrates cases for  $\nu=3, \ 10, \ 50$  as well as that for an initial
uniform density.

Our approach has considered the determinate rate process for the evolution of 
individual types with an inhomogeneous fitness distribution; Baake and Georgii~\cite{Baake} consider 
the multi-dimensional systems of many types. Their mutation and differential reproduction processes 
yield n-dimensional systems of differential equations which they analyse using variational methods.

\end{document}